\documentclass[onecolumn]{pasj00}
\SetRunningHead{Hosono, Saitoh \& Makino}{Density Independent SPH for non-ideal EOS}
\Received{}
\Accepted{2013/07/02}
\Published{}
\begin{document}
\title{Density Independent Smoothed Particle Hydrodynamics for Non-Ideal Equation of State}
\author{Natsuki \textsc{Hosono}\altaffilmark{1}, Takayuki R. \textsc{Saitoh}\altaffilmark{2} and Junichiro \textsc{Makino}\altaffilmark{2,3}}
\altaffiltext{1}{Department of Earth \& Planetary Sciences, Tokyo Institute of Technology, Ookayama, Meguro-ku, Tokyo 152-8551}
\altaffiltext{2}{Earth-Life Science Institute, Tokyo Institute of Technology, Ookayama, Meguro-ku, Tokyo 152-8550}
\altaffiltext{3}{RIKEN Advanced Institute for Computational Science, Minatojima-minamimachi, Chuo-ku, Kobe, Hyogo 650-0047}
\email{natsuki.h@geo.titech.ac.jp}
\maketitle

\begin{abstract}
The smoothed particle hydrodynamics (SPH) method is a useful numerical tool for the study of a variety of astrophysical and planetlogical problems.
However, it turned out that the standard SPH algorithm has problems in dealing with hydrodynamical instabilities.
This problem is due to the assumption that the local density distribution is differentiable.
In order to solve this problem, a new SPH formulation, which does not require the differentiability of the density, have been proposed.
This new SPH method improved the treatment of hydrodynamical instabilities.
This method, however, is applicable only to the equation of state (EOS) of the ideal gas.
In this paper, we describe how to extend the new SPH method to non-ideal EOS.
We present the results of various standard numerical tests for non-ideal EOS.
Our new method works well for non-ideal EOS.
We conclude that our new SPH can handle hydrodynamical instabilities for an arbitrary EOS and that it is an attractive alternative to the standard SPH.
\end{abstract}

\section{Introduction}
In the field of astrophysics and planetary science, fluid dynamical processes play important roles on virtually all length and mass scales from galaxies to planets.
The smoothed particle hydrodynamics (SPH) method (Lucy 1977; Gingold \& Monaghan 1977) is one of the most popular simulation methods to solve the motion of a fluid in astrophysical problems (for reviews, see Monaghan 1992; Rosswog 2009; Springel 2010).
In the SPH method, fluid elements are represented by hypothetical particles (so-called SPH particles).
Thus, the dynamical equations are written in the Lagrangian form of hydrodynamical equations.
Compared to the grid-base methods, the SPH method is suitable to problems in which inhomogeneities such as large empty regions and small dense core develop.
Furthermore, it is easy to incorporate various physical effects to the SPH scheme, such as self-gravity, radiative cooling and chemical reactions.
Because of these advantages, various astrophysical problems, such as star formation, planetesimal collisions and galaxy formation, have been studied using the SPH method.

Recently, however, it has been pointed out that the standard SPH method has difficulties in dealing with hydrodynamical instabilities, such as Kelvin-Helmholtz instability (KHI) or Rayleigh-Taylor instability (RTI) (e.g., Okamoto et al. 2003; Agertz et al. 2007; Valcke et al. 2010; McNally et al. 2012).
Agertz et al. (2007) has concluded that this difficulty is due to the requirement of the standard SPH that the density must be continuous and differentiable.
This requirement is not satisfied at contact discontinuities.
As a result, at contact discontinuities, the pressure of the low-density side is overestimated and that of the high-density side is underestimated.
Thus, pressure is also overestimated at the low-density side of the contact discontinuity and ``unphysical" repulsive force appears.
This unphysical repulsive force causes a surface tension effect which suppresses the growth of hydrodynamical instabilities.

To resolve this issue, modifications of the standard SPH method have been proposed.
Price (2008) introduced the artificial thermal conductivity term in the SPH equation to smooth the thermal energy at the contact discontinuity (Price 2008; Valdarnini 2012).
Cha et al. (2010) and Murante et al. (2011) showed that the Godunov SPH, originally developed by Inutsuka (2002), can describe hydrodynamical instabilities.
However, the Godunov scheme is difficult to extend to non-ideal EOS, though methods exist (e.g., Colella \& Glaz 1985).
Read et al. (2010) showed that KHI takes place with the equation of motion of Ritchie \& Thomas (2001) and with a kernel function which has larger number of neighbours.
He\ss \, \& Springel (2010) replaces the density estimate in the standard SPH by a new density estimate with Voronoi tessellation.
Abel (2011) used the relative pressure instead of the absolute value of the pressure in the equation of motion.
However, this approach does not satisfy the conservation of momentum.
Garc\'ia-Senz et al. (2012) present a new formulation which is based on a tensor approach.

Saitoh \& Makino (2013) have proposed a new formulation of SPH.
They pointed out that the problematic requirement of the differentiability of the density arises from the formula used to estimate the volume element associated with a particle in the standard SPH.
The volume element used in the standard SPH is $\Delta V_i = m_i / \rho_i$, where $\Delta V_i$, $m_i$ and $\rho_i$ are the volume element, the mass and the density of a particle $i$, respectively.
Thus, by using an estimate of the volume element which is independent of the mass and density, we can avoid the necessity for the differentiability of the density.
In particular, Saitoh \& Makino (2013) used $U_i / q_i$ as the volume element, where $U_i$ is the internal energy and $q_i$ is the energy spatial density.
As a result, their formulation does not require the differentiability of the density.
In the case of the equation of state (EOS) of ideal gas, the pressure is proportional to the energy spatial density.
Thus, the requirement of the differentiability of the energy spatial density corresponds to the requirement of the differentiability of the pressure.
Their formulation does not introduce any physically non-existent term and does not break any conservation property.
They showed that their new SPH can handle hydrodynamical instabilities well.

However, their new formulation can be applied only to ideal gas.
In many astrophysical problems, the EOS is non-ideal.
In this paper, we present an extension of their new SPH to non-ideal EOS.
As is shown in Saitoh \& Makino (2013), we can choose an arbitrary basis for the volume element.
Thus, using a different choice of the estimate of the volume element, we can extend the new SPH to non-ideal EOS, without losing any advantages of Saitoh \& Makino (2013)'s formulation.

This paper is organized as follows.
In \S 2, we present a brief overview of the formulation of Saitoh \& Makino (2013).
Then in \S 3, our new formulation is described.
In \S 4, the results of various test calculations with our new SPH are shown.
Finally in \S 5, we summarize this paper.

\section{Overview of the Density Independent SPH}
Let us consider the following set of equations that describes the motion of fluid:
\begin{eqnarray}
\frac{d \rho}{dt} & = & - \rho \nabla \cdot \boldsymbol{v}, \label{eq:continuity}\\
\frac{d \boldsymbol{v}}{dt} & = & - \frac{1}{\rho}\nabla p, \label{eq:motion}\\
\frac{d u}{dt} & = & -\frac{p}{\rho} \nabla \cdot \boldsymbol{v}. \label{eq:energy}
\end{eqnarray}
Here, $\rho$, $\boldsymbol{v}$, $p$, $u$ and $t$ are density, velocity, pressure, specific internal energy and time, respectively.
The pressure $p$ is given by EOS, $p = p(\rho , u)$.

In the SPH method, a fluid is expressed by a number of SPH particles.
Physical quantities at a point are approximated by the summation of the contributions of these particles.
First, we approximate a function $f(\boldsymbol{x})$ by the convolution with a kernel function $W(\boldsymbol{x}, h)$:
\begin{eqnarray}
\langle f \rangle (\boldsymbol{x}) = \int f(\boldsymbol{x}') W(\boldsymbol{x} - \boldsymbol{x}'; h) dV', \label{eq:conv}
\end{eqnarray}
where $\boldsymbol{x}$ is the position vector, $W$ is the kernel function and $h$ is the smoothing length.
The kernel function must be differentiable for $|\boldsymbol{x}|$ and have following two properties:
\begin{eqnarray}
\int W(\boldsymbol{x} - \boldsymbol{x}'; h) dV' & = & 1,\\
\lim_{h \rightarrow 0} W(\boldsymbol{x} - \boldsymbol{x}'; h) & = & \delta(\boldsymbol{x} - \boldsymbol{x}').
\end{eqnarray}
We can use an arbitrary kernel function, as long as the above conditions are satisfied.
Throughout this paper, we use the cubic spline function proposed by Monaghan \& Lattanzio (1985).
\begin{eqnarray}
W(\boldsymbol{x}; h) = \frac{\sigma}{h^D}
\left\{ \begin{array}{ll}
\displaystyle \frac{1}{4}(4 - 6 s^2 + 3 s^3) & \left(0 \leqq s < 1\right), \\[2ex]
\displaystyle \frac{1}{4}\left( 2 - s \right)^3 & \left(1 \leqq  s < 2\right), \\[2ex]
0 & \left(2 \leqq s\right), \\
\end{array} \right.
\end{eqnarray}
where $s = |\boldsymbol{x}| / h$, $D$ is number of dimensions and $\sigma$ is the normalization constant that takes the value of $2/3, 10/7\pi, 1/\pi$ in one-, two- and three-dimensional cases, respectively.
Note that the use of this cubic spline kernel for the derivative sometimes causes clustering of the SPH particles.
In order to avoid this clustering of the SPH particles, we adopt a gradient of the kernel which has a triangular shape, as Thomas \& Couchman (1992) did:
\begin{eqnarray}
\nabla W(\boldsymbol{x}; h) = - \frac{\sigma \boldsymbol{x}}{h^{D + 1} |\boldsymbol{x}|}
\left\{ \begin{array}{ll}
1 & \left(0 \leqq s < 2/3\right), \\
\displaystyle \frac{3}{4}s(4 - 3s) & \left(2/3 \leqq s < 1\right), \\[2ex]
\displaystyle \frac{3}{4}\left( 2 - s \right)^2 & \left(1 \leqq  s < 2\right), \\[2ex]
0 & \left(2 \leqq s \right). \\
\end{array} \right.
\end{eqnarray}
The essential solution is to adopt the kernels which do not show the pairing instability (Read et al. 2010; Dehnen \& Aly 2012).

In order to evaluate the value of the physical quantities at positions of particles, we need to discretize Eq. ($\ref{eq:conv}$).
By approximating the integral by summation over particles, we obtain the following equation:
\begin{eqnarray}
\langle f \rangle (\boldsymbol{x}) = \sum_{j} f_j W(\boldsymbol{x} - \boldsymbol{x}_j ; h) \Delta V_j, \label{eq:smth}
\end{eqnarray}
where the subscript $j$ denotes particle index and $f_j$ is the value of $f(\boldsymbol{x})$ of particle $j$.
In the formulation of Saitoh \& Makino (2013), the volume element $\Delta V_j$ is replaced by $U_j / q_j$, where $U_j = m_j u_j$ is the internal energy and $q_j = \rho_j u_j$ is the energy spatial density of the $j$-th particle.
Thus, $f_i$ can be written as follows:
\begin{eqnarray}
f_i = \sum_{j} f_j \frac{U_j}{q_j} W(\boldsymbol{x}_{ij} ; h), \label{eq:smth_f_i}
\end{eqnarray}
where $\boldsymbol{x}_{ij} = \boldsymbol{x}_i - \boldsymbol{x}_j$.
By substituting $q$ into $f$, we obtain
\begin{eqnarray}
q_i = \sum_{j} U_j W(\boldsymbol{x}_{ij} ; h). \label{eq:di_q}
\end{eqnarray}
From Eq. ($\ref{eq:smth_f_i}$), the derivative of $f_i$ is given by
\begin{eqnarray}
\nabla f_i = \sum_{j} f_j \frac{U_j}{q_j} \nabla W(\boldsymbol{x}_{ij} ; h). \label{eq:nabla_smth}
\end{eqnarray}
Now we first derive the equation of energy and then the equation of motion.

In order to derive the equation of energy, we need an expression of $\nabla \cdot \boldsymbol{v}$.
We use the following relation:
\begin{eqnarray}
\nabla(q\boldsymbol{v}) = \boldsymbol{v} \cdot \nabla q + q \nabla \cdot \boldsymbol{v}. \label{eq:di_nabla_v}
\end{eqnarray}
Note that in the case of the ideal gas, the pressure is proportional to $q$.
Thus, around the contact discontinuity, $q$ is differentiable.
By applying Eq. ($\ref{eq:nabla_smth}$) to Eq. ($\ref{eq:di_nabla_v}$), we obtain
\begin{eqnarray}
q_i \nabla \cdot \boldsymbol{v}_i = - \sum_j U_j \boldsymbol{v}_{ij} \cdot \nabla W(\boldsymbol{x}_{ij}; h), \label{eq:di_nabla_v2}
\end{eqnarray}
where $\boldsymbol{v}_{ij} = \boldsymbol{v}_i - \boldsymbol{v}_j$.
Here, the density $\rho_i$ is
\begin{eqnarray}
\rho_i = \frac{m_i q_i}{U_i}. \label{eq:di_density}
\end{eqnarray}
By applying Eq. ($\ref{eq:di_nabla_v2}$) and Eq. ($\ref{eq:di_density}$) to Eq. ($\ref{eq:energy}$), we can write the equation of energy as
\begin{eqnarray}
\frac{dU_i}{dt} = \sum_j \frac{U_i p_i}{q_i^2} U_j \boldsymbol{v}_{ij} \cdot \nabla W(\boldsymbol{r}_{ij}; h). \label{eq:di_energy}
\end{eqnarray}
Now we define the change in the internal energy of the $i$-th particle due to the interaction with the $j$-th particle as $dU_{ij} / dt$.
From Eq. ($\ref{eq:di_energy}$), we obtain
\begin{eqnarray}
\frac{dU_{ij}}{dt} = \frac{U_i U_j p_i}{q_i^2} \boldsymbol{v}_{ij} \cdot \nabla W(\boldsymbol{x}_{ij} ; h). \label{eq:d_U_ij}
\end{eqnarray}

From the equation of energy, we derive the equation of motion.
The change of the internal energy is the same as that of the kinetic energy with an opposite sign;
\begin{eqnarray}
\frac{dU_{ij}}{dt} + \frac{dU_{ji}}{dt} = - \frac{d}{dt} (K_i + K_j), \label{eq:K_eq_U}
\end{eqnarray}
where $K_i$ and $K_j$ are the kinetic energy of the $i$-th and $j$-th particle, respectively.
Here we consider the change of $K_i$ due to the interaction with the $j$-th particle only.
From Eq. ($\ref{eq:d_U_ij}$), the left hand side of Eq. ($\ref{eq:K_eq_U}$) can be written as
\begin{eqnarray}
\frac{dU_{ij}}{dt} + \frac{dU_{ji}}{dt} = U_i U_j \left( \frac{p_i}{q^2_i} + \frac{p_j}{q^2_j}\right) \boldsymbol{v}_{ij} \cdot \nabla W(\boldsymbol{x}_{ij} ; h). \label{eq:U_ij}
\end{eqnarray}
Here, $K_i + K_j$ is
\begin{eqnarray}
K_i + K_j & = & \frac{1}{2} m_i \boldsymbol{v}_i^2 + \frac{1}{2} m_j \boldsymbol{v}_j^2, \nonumber\\
& = & \frac{1}{2}(m_i + m_j)\left( \frac{m_i \boldsymbol{v}_i + m_j \boldsymbol{v}_j}{m_i + m_j} \right)^2 + \frac{1}{2}\frac{m_i m_j}{m_i + m_j} \boldsymbol{v}_{ij}^2.
\end{eqnarray}
Thus, the change of the kinetic energy can be written as
\begin{eqnarray}
\frac{d}{dt} (K_i + K_j) = \frac{1}{2}(m_i + m_j)\frac{d}{dt}\left( \frac{m_i \boldsymbol{v}_i + m_j \boldsymbol{v}_j}{m_i + m_j} \right)^2 + \frac{m_i m_j}{m_i + m_j} \boldsymbol{v}_{ij} \cdot \frac{d\boldsymbol{v}_{ij}}{dt}. \label{eq:kinetic_energy}
\end{eqnarray}
Since the total momentum of two particles is conserved, we have
\begin{eqnarray}
\frac{d}{dt}(m_i \boldsymbol{v}_i + m_j \boldsymbol{v}_j) = 0. \label{eq:momentum}
\end{eqnarray}
Thus, the first term of the right hand side of Eq. ($\ref{eq:kinetic_energy}$) is zero.
By substituting Eqs. ($\ref{eq:U_ij}$) and ($\ref{eq:kinetic_energy}$) into Eq. ($\ref{eq:K_eq_U}$), we obtain
\begin{eqnarray}
- \frac{m_i m_j}{m_i + m_j} \frac{d\boldsymbol{v}_{ij}}{dt} = U_i U_j \left( \frac{p_i}{q^2_i} + \frac{p_j}{q^2_j} \right) \nabla W(\boldsymbol{x}_{ij} ; h). \label{eq:di_motion1}
\end{eqnarray}
By using Eq. ($\ref{eq:momentum}$), we can eliminate $\boldsymbol{v}_j$ in Eq. ($\ref{eq:di_motion1}$) and we finally obtain
\begin{eqnarray}
m_i \frac{d\boldsymbol{v}_i}{dt} = - \sum_{j} U_i U_j \left( \frac{p_i}{q^2_i} + \frac{p_j}{q^2_j} \right)\nabla W(\boldsymbol{x}_{ij} ; h). \label{eq:di_motion}
\end{eqnarray}
Note that for the case of a variable kernel size, $\nabla W(\boldsymbol{x}_{ij}; h)$ must take symmetrical form in the smoothing length to satisfy the conservation of energy and momentum.
This condition is achieved by replacing $\nabla W(\boldsymbol{x}_{ij}; h)$ with $[\nabla W(\boldsymbol{x}_{ij}; h_i) + \nabla W(\boldsymbol{x}_{ij}; h_j)] / 2$ or $\nabla W[\boldsymbol{x}_{ij}; (h_i + h_j) / 2]$.
Throughout this paper, we adopt the former form.

Hopkins (2013) and Saitoh \& Makino (2013) have derived the equation of motion from a Lagrangian.
The advantage of this derivation is that it includes the variation of $h$ naturally.
This term, so-called the ``$\nabla h$" term, is important in simulations in which extremely strong shocks present (see section 3.5 in Saitoh \& Makino, 2013).

\section{Extension to non-ideal EOS of DISPH}
In the previous section, we summarized the formulation of Saitoh \& Makino (2013).
As stated above, the formulation of Saitoh \& Makino (2013) has one assumption that the pressure is proportional to the energy spatial density.
In this section, we extend their new SPH to an arbitrary EOS.
In order to construct a new SPH formulation, we introduce the following quantity:
\begin{eqnarray}
Y_i = p_i \Delta V_i.
\end{eqnarray}
In our formulation, we use the following new volume element:
\begin{eqnarray}
\Delta V_i = \frac{Y_i}{p_i}. \label{eq:new_vol}
\end{eqnarray}
By substituting Eq. $(\ref{eq:new_vol})$ into Eq. $(\ref{eq:smth})$, we obtain the following two equations:
\begin{eqnarray}
f_i & = & \sum_{j} f_j  \frac{Y_j}{p_j} W(\boldsymbol{x}_{ij} ; h), \label{eq:pf_f}\\
\nabla f_i & = & \sum_{j} f_j \frac{Y_j}{p_j} \nabla W(\boldsymbol{x}_{ij} ; h). \label{eq:pf_nabla_f}
\end{eqnarray}
By substituting $p$ into $f$ in Eq. ($\ref{eq:pf_f}$), we obtain the smoothed pressure $p$,
\begin{eqnarray}
p_i = \sum_{j} Y_j W(\boldsymbol{x}_{ij} ; h) \label{eq:p}.
\end{eqnarray}

We first derive the equation of energy, and then we derive the equation of motion, following the derivation of Saitoh \& Makino (2013).

\subsection{Equation of Energy}
In order to derive the equation of energy, we need the expression of $\nabla \cdot \boldsymbol{v}$ in our new SPH.
Here, we use following relation:
\begin{eqnarray}
p \nabla \cdot \boldsymbol{v} = \nabla \cdot (p \boldsymbol{v}) - \boldsymbol{v} \cdot \nabla p, \label{eq:p_v1}
\end{eqnarray}
which can be obtained by replacing $q$ in Eq. $(\ref{eq:di_nabla_v})$ by $p$.
Thus, the expression of $\nabla \cdot \boldsymbol{v}$ is
\begin{eqnarray}
\nabla \cdot \boldsymbol{v}_i = - \frac{1}{p_i} \sum_j Y_j \boldsymbol{v}_{ij} \cdot \nabla W(\boldsymbol{x}_{ij}; h) \label{eq:pf_div_v}.
\end{eqnarray}
In our new SPH, the density $\rho_i$ can be expressed as
\begin{eqnarray}
\rho_i = \frac{m_i p_i}{Y_i}. \label{eq:pf_density}
\end{eqnarray}
By applying Eqs. ($\ref{eq:pf_div_v}$) and ($\ref{eq:pf_density}$) to Eq. ($\ref{eq:energy}$), the equation of energy can be written as
\begin{eqnarray}
\frac{dU_i}{dt} = \sum_j \frac{Y_i Y_j}{p_i} \boldsymbol{v}_{ij} \cdot \nabla W(\boldsymbol{x}_{ij}; h). \label{eq:pf_energy}
\end{eqnarray}
Hence the equation corresponding to Eq. $(\ref{eq:d_U_ij})$ is
\begin{eqnarray}
\frac{dU_{ij}}{dt} = \frac{Y_i Y_j}{p_i} \boldsymbol{v}_{ij} \cdot \nabla W(\boldsymbol{x}_{ij}; h). \label{eq:pf_U_ij}
\end{eqnarray}

\subsection{Equation of Motion}
From the equation of energy, we derive the equation of motion.
By using Eq. ($\ref{eq:pf_U_ij}$) instead of Eq. ($\ref{eq:d_U_ij}$), we obtain the analogue of Eq. ($\ref{eq:di_motion1}$):
\begin{eqnarray}
- \frac{m_i m_j}{m_i + m_j} \frac{d\boldsymbol{v}_{ij}}{dt} = Y_i Y_j \left( \frac{1}{p_i} + \frac{1}{p_j} \right) \nabla W(\boldsymbol{x}_{ij}; h). \label{eq:pf_motion1}
\end{eqnarray}
Thus the equation of motion becomes
\begin{eqnarray}
m_i \frac{d\boldsymbol{v}_i}{dt} = - \sum_{j} Y_i Y_j \left( \frac{1}{p_i} + \frac{1}{p_j} \right)\nabla W(\boldsymbol{x}_{ij}; h). \label{eq:pf_motion}
\end{eqnarray}

\subsection{The equation for $Y$}
In the previous section, we derived the equation of energy and the equation of motion.
These equations determine the evolution of fluid.
However, in order to actually perform the numerical integration, we need to determine new values of pressure, by solving implicit equation, Eq. ($\ref{eq:p}$), for  a given position $\boldsymbol{x}_i$ and specific internal energy $u_i$.
We solve Eq. ($\ref{eq:p}$) by iteration.
Here we summarize the actual procedure.\\
\textbf{Step1}: We calculate the density using Eq. ($\ref{eq:pf_density}$).\\
\textbf{Step2}: From the EOS, density and internal energy, we obtain the non-smoothed pressure $\hat{p} = p(\rho, u)$.\\
\textbf{Step3}: We update $\hat{Y}$ from the equation $\hat{Y} = m \hat{p} / \rho$.\\
\textbf{Step4}: We calculate $p$ by using Eq. ($\ref{eq:p}$). If necessary, we go back to Step1.\\
Unless otherwise noted, only one cycle of the above iteration is applied.

We calculate the initial guess of $\hat{Y}$ by integrating the time derivative of $Y$.
It is given by
\begin{eqnarray}
\frac{dY}{dt} & = & p \frac{d \Delta V}{d t} + \Delta V \frac{d p}{d t}, \nonumber\\
& = & p \Delta V \left( \frac{1}{\Delta V} \frac{d \Delta V}{d t} + \frac{1}{p} \frac{d p}{d t} \right), \nonumber\\
& = & Y \left( \frac{1}{\Delta V} \frac{d \Delta V}{d t} + \frac{\Delta V}{p}\frac{1}{\Delta V}\frac{d \Delta V}{d t}\frac{\partial p}{\partial \Delta V} \right), \nonumber\\
& = & Y \frac{1}{\Delta V}\frac{d \Delta V}{d t} \left( 1 + \frac{\Delta V}{p}\frac{\partial p}{\partial \Delta V} \right). \label{eq:dY_dt}
\end{eqnarray}
From the continuity equation, Eq. ($\ref{eq:continuity}$), and $\Delta V = m / \rho$, we obtain
\begin{eqnarray}
\frac{1}{\Delta V}\frac{d \Delta V}{dt} & = & \nabla \cdot \boldsymbol{v}, \label{eq:pf_continuity}\\
\frac{\Delta V}{p} \frac{\partial p}{\partial \Delta V} & = & - \frac{\rho}{p}\frac{\partial p}{\partial \rho}. \label{eq:pf_dp_dV}
\end{eqnarray}
By substituting Eqs. ($\ref{eq:pf_continuity}$) and ($\ref{eq:pf_dp_dV}$) into Eq. ($\ref{eq:dY_dt}$), we obtain
\begin{eqnarray}
\frac{dY_i}{dt} = (\gamma_i - 1)\sum_j \frac{Y_i Y_j}{p_i} \boldsymbol{v}_{ij} \cdot \nabla W(\boldsymbol{x}_{ij} ; h) \label{eq:eq_Y},
\end{eqnarray}
where
\begin{eqnarray}
\gamma_i := \frac{\rho_i}{p_i}\left(\frac{\partial p}{\partial \rho}\right)_i.
\end{eqnarray}

When Eq. ($\ref{eq:p}$) is satisfied, we should have $p_i = p(\rho_i, u_i)$.
If we express this as equation for specific internal energy $u_i$, we have $u (p_i, \rho_i) = u_i$, where $u (p_i, \rho_i)$ comes from the EOS.
Therefore, the quantity
\begin{eqnarray}
\Delta u = \frac{1}{N} \sum_i \Delta u_i, \label{eq:err}
\end{eqnarray}
gives the accuracy at which Eq. ($\ref{eq:p}$) is satisfied, where $N$ is the total number of the SPH particles and
\begin{eqnarray}
\Delta u_i = \frac{|u(\rho_i, p_i) - u_i|}{u_i}. \label{eq:err2}
\end{eqnarray}

\subsection{Smoothing length}
The smoothing length is the effective length of the kernel function.
In general, individual and time-varying smoothing length are used.
In this paper, we use the following equation to determine $h_i$:
\begin{eqnarray}
h_i = \eta \left( \frac{m_i}{\rho_i} \right)^{1/D} = \eta \left( \frac{Y_i}{p_i} \right)^{1/D}.
\end{eqnarray}
Unless otherwise specified, we set the parameter $\eta = 1.2$.

\subsection{A conservative formulation of our new SPH using the action principle}
As stated above, we derived the equation of motion and equation of energy of our new SPH from the fundamental equations of fluid.
However, as Springel \& Hernquist (2002) did, the equations for the SPH method can be also derived from the Lagrangian.
Recently, Hopkins (2013) derived the equation of motion for Saitoh \& Makino (2013)'s new SPH from the Lagrangian.
In this section, we derive the equation of motion for our new SPH from the Lagrangian and show the corresponding expression of $\nabla h$ term for our new SPH.

Here we consider the Euler-Lagrange equation:
\begin{eqnarray}
\frac{d}{dt}\frac{\partial L}{\partial \dot{Q}_i} - \frac{\partial L}{\partial Q_i} = \sum_j \lambda_j \frac{\partial \phi_j}{\partial Q_i}, \label{eq:EL}
\end{eqnarray}
where $L$, $\lambda_i$ and $\phi_i$ are Lagrangian, Lagrange multipliers and appropriate constraints, respectively.
According to Hopkins (2013), we use the following constraint equation:
\begin{eqnarray}
\phi_i = \frac{4 \pi}{3} (H h_i)^3 \frac{1}{\Delta V_i} - N_{\rm SPH} = 0, \label{eq:phi}
\end{eqnarray}
where $H$ is the kernel-support radius.
This constraint equation gives a condition that there are an approximately constant number of particle in the kernel for three dimensions, if the mass of each SPH particle is equal.
The Lagrangian can be written as follows:
\begin{eqnarray}
L(\boldsymbol{Q}, \dot{\boldsymbol{Q}}) = \sum_j \left(\frac{1}{2} m_j \boldsymbol{v}_j^2 - m_j u_j\right), \label{eq:L}
\end{eqnarray}
where $\boldsymbol{Q} = (\boldsymbol{x}_1, ..., \boldsymbol{x}_N, h_1, ..., h_N)$.
By substituting Eqs. (\ref{eq:phi}) and (\ref{eq:L}) into Eq. (\ref{eq:EL}), we obtain $2N$ equations.

Let us consider the second half of the above $2N$ equations.
By substituting $h_i$ into $Q_i$, the right-hand side of the Euler-Lagrange equation becomes
\begin{eqnarray}
\sum_j \lambda_j \frac{\partial \phi_j}{\partial h_i} & = & \lambda_i \frac{\partial }{\partial h_i} \left[ \frac{4 \pi}{3} (H h_i)^3 \frac{1}{\Delta V_i} - N_{\rm SPH}\right], \nonumber \\
& = & \lambda_i \frac{4 \pi H^3 h_i^2}{\Delta V_i} \left(1 - \frac{h_i}{3 \Delta V_i} \frac{\partial \Delta V_i}{\partial h_i}\right) \label{eq:rhs}.
\end{eqnarray}
The left-hand side becomes
\begin{eqnarray}
\frac{d}{dt}\frac{\partial L}{\partial \dot{h}_i} - \frac{\partial L}{\partial h_i} & = & \frac{\partial (m_i u_i)}{\partial h_i}, \nonumber\\
& = & - p_i \frac{\partial \Delta V_i}{\partial h_i} \label{eq:lhs}.
\end{eqnarray}
Note that here we used the following relation from the first law of thermodynamics:
\begin{eqnarray}
\frac{\partial U_i}{\partial Q_i} = - p_i \frac{\partial \Delta V_i}{\partial Q_i}. \label{eq:1st_law}
\end{eqnarray}
From Eqs. (\ref{eq:rhs}) and (\ref{eq:lhs}), we obtain the Lagrangian multipliers as follows:
\begin{eqnarray}
\lambda_i = - \frac{p_i \Delta V_i}{4 \pi H^3 h_i^2} \frac{\partial \Delta V_i}{\partial h_i} \left(1 - \frac{h_i}{3 \Delta V_i} \frac{\partial \Delta V_i}{\partial h_i} \right)^{-1} \label{eq:lambda}.
\end{eqnarray}

By substituting the positions of SPH particles into $Q_i$ in Eqs. (\ref{eq:EL}) and (\ref{eq:lambda}), we obtain equation of motion:
\begin{eqnarray}
m_i \frac{d \boldsymbol{v}_i}{dt} = \sum_j \left(p_j - \frac{4}{3}\pi H^3 h_j^3 \lambda_j \frac{1}{\Delta V_j^2}\right) \nabla (\Delta V_j) \label{eq:Lag_EOM}.
\end{eqnarray}

Here we recall that in our new SPH, the volume element $\Delta V_i$ is estimated as $Y_i / p_i$.
Thus we obtain following equations:
\begin{eqnarray}
\frac{\partial \Delta V_i}{\partial h_i} & = & - \frac{Y_i}{p_i^2} \frac{\partial p_i}{\partial h_i}, \label{eq:pp_ph}\\
\nabla (\Delta V_j) & = & - \frac{Y_j}{p_j^2} \nabla p_j, \nonumber \\
& = & - \frac{Y_j}{p_j^2} \left[ Y_i \nabla W(\boldsymbol{x}_{ij}; h_j) + \delta_{ij} \sum_k Y_k \nabla W(\boldsymbol{x}_{ik}; h_i) \right] \label{eq:nabla_DV},
\end{eqnarray}
where $\delta_{ij}$ is Kronecker's delta.
By substituting Eq. (\ref{eq:pp_ph}) into Eq. (\ref{eq:lambda}), we obtain $\lambda_i$:
\begin{eqnarray}
\lambda_i = \frac{Y_i^2}{4 \pi H^3 h_i^2 p_i^2} \frac{\partial p_i}{\partial h_i}\left( 1 + \frac{h_i}{3 p_i} \frac{\partial p_i}{\partial h_i} \right)^{-1} \label{eq:pf_lambda}.
\end{eqnarray}
By substituting Eqs. (\ref{eq:nabla_DV}) and (\ref{eq:pf_lambda}) into Eq. ($\ref{eq:Lag_EOM}$), we obtain the equation of motion as follows:
\begin{eqnarray}
m_i \frac{d \boldsymbol{v}_i}{dt} = - \sum_j Y_i Y_j \left[ \frac{f_{i}^{\rm grad}}{p_i} \nabla W(\boldsymbol{x}_{ij}; h_i) + \frac{f_{j}^{\rm grad}}{p_j} \nabla W(\boldsymbol{x}_{ij}; h_j) \right], \label{eq:EoM_h}
\end{eqnarray}
where
\begin{eqnarray}
f_{i}^{\rm grad} = \left(1 + \frac{h_i}{3 p_i} \frac{\partial p_i}{\partial h_i}\right)^{-1}.
\end{eqnarray}

In order to calculate the time evolution of the specific internal energy explicitly, we need the equation of energy.
We derive the equation of energy with $\nabla h$ term as follows.
From the first law of thermodynamics, we obtain
\begin{eqnarray}
\frac{dU_i}{dt} & = & - p_i \frac{d \Delta V_i}{d t} \label{eq:dU_dt}.
\end{eqnarray}
Here, 
\begin{eqnarray}
\frac{d \Delta V_i}{d t} & = & \frac{d}{dt}\left( \frac{Y_i}{p_i} \right), \nonumber\\
& = & - \frac{Y_i}{p_i^2} \frac{d p_i}{dt}, \nonumber\\
& = & - \frac{Y_i}{p_i^2} \frac{d}{dt} \sum_j Y_j W(\boldsymbol{x}_{ij}; h_i), \nonumber\\
& = & - \frac{Y_i}{p_i^2} \sum_j Y_j \left[ \boldsymbol{v}_{ij} \cdot \nabla W(\boldsymbol{x}_{ij}; h_i) + \frac{dh_i}{dt} \frac{\partial W(\boldsymbol{x}_{ij}; h_i)}{\partial h_i} \right] \label{eq:DeltaV_dt1}.
\end{eqnarray}
From Eq. (\ref{eq:phi}), we obtain following equation:
\begin{eqnarray}
\frac{dh_i}{dt} = \frac{h_i}{3 \Delta V_i}\frac{d \Delta V_i}{dt} \label{eq:dh_dt}.
\end{eqnarray}
By substituting Eq. (\ref{eq:dh_dt}) into Eq. (\ref{eq:DeltaV_dt1}),
\begin{eqnarray}
\frac{d \Delta V_i}{d t} & = & - \frac{Y_i}{p_i^2} \sum_j Y_j \left[ \boldsymbol{v}_{ij} \cdot \nabla W(\boldsymbol{x}_{ij}; h_i) + \frac{h_i}{3 \Delta V_i} \frac{\partial W(\boldsymbol{x}_{ij}; h_i)}{\partial h_i} \frac{d \Delta V_i}{dt} \right] \label{eq:DeltaV_dt}, \nonumber \\
& = & - \frac{Y_i}{p_i^2} \sum_j Y_j \boldsymbol{v}_{ij} \cdot \nabla W(\boldsymbol{x}_{ij}; h_i) - \frac{h_i}{3 p_i} \frac{\partial p_i}{\partial h_i} \frac{d \Delta V_i}{dt}.
\end{eqnarray}
From the above equation we obtain
\begin{eqnarray}
\frac{d \Delta V_i}{dt} = - f_{i}^{\rm grad} \frac{Y_i}{p_i^2} \sum_j Y_j \boldsymbol{v}_{ij} \cdot \nabla W(\boldsymbol{x}_{ij}; h_i). \label{eq:DeltaV_dt2}
\end{eqnarray}
By substituting Eq. (\ref{eq:DeltaV_dt2}) into Eq. (\ref{eq:dU_dt}), we obtain the equation of energy as follows:
\begin{eqnarray}
\frac{dU_i}{dt} = \sum_j \frac{Y_i Y_j}{p_i} f_{i}^{\rm grad} \boldsymbol{v}_{ij} \cdot \nabla W(\boldsymbol{x}_{ij}; h_i).
\end{eqnarray}
With the equation of energy, we can obtain the analogue to Eq. $(\ref{eq:eq_Y})$ as follows:
\begin{eqnarray}
\frac{dY_i}{dt} = (\gamma_i - 1) \sum_j \frac{Y_i Y_j}{p_i} f_{i}^{\rm grad} \boldsymbol{v}_{ij} \cdot \nabla W(\boldsymbol{x}_{ij}; h_i).
\end{eqnarray}

Note that here we discuss only the three-dimensional case.
However, by using appropriate constraint, we can easily derive the expression of $f_{i}^{\rm grad}$ in one- or two-dimensional case as follows:
\begin{eqnarray}
f_{i}^{\rm grad} = \left(1 + \frac{1}{D}\frac{h_i}{p_i} \frac{\partial p_i}{\partial h_i}\right)^{-1}. \label{eq:f_grad}
\end{eqnarray}

\subsection{Artificial Viscosity}
We need to introduce artificial viscosity to handle shocks.
There are several different forms of artificial viscosity (e.g., Lattanzio \& Monaghan 1985; Monaghan 1997).
In this paper we adopt the following form of the artificial viscosity proposed by Monaghan (1997).
It is expressed as
\begin{eqnarray}
\Pi_{ij} = - \displaystyle\frac{\alpha_{\rm AV}}{2} \displaystyle\frac{v_{ij}^{\rm sig}w_{ij}}{\rho_{ij}},
\end{eqnarray}
where
\begin{eqnarray}
v_{ij}^{\rm sig} & = & c_i + c_j - 3 w_{ij},\\
w_{ij} & = & \left\{ \begin{array}{cl}
\displaystyle\frac{\boldsymbol{v}_{ij} \cdot \boldsymbol{x}_{ij}}{| \boldsymbol{x}_{ij} |} & \quad {\rm if} \quad \boldsymbol{x}_{ij} \cdot \boldsymbol{v}_{ij} < 0,\\[2ex]
0 & \quad {\rm otherwise},
\end{array} \right.\\
\rho_{ij} & = & \frac{1}{2}(\rho_i + \rho_j) \label{eq:rho_ij}.
\end{eqnarray}
Note that the use of Eq. ($\ref{eq:pf_density}$) for the calculation of the artificial viscosity sometimes leads to unstable behaviour under strong shocks.
It seems to be safer to use the smoothed density,
\begin{eqnarray}
\rho_i = \sum_{j} m_j W(\boldsymbol{x}_{ij} ; h_i).
\end{eqnarray}

In order to suppress the shear viscosity, we apply the Balsara switch (Balsara 1995).
It is given by
\begin{eqnarray}
F_i = \frac{| \nabla \cdot \boldsymbol{v}_i |}{| \nabla \cdot \boldsymbol{v}_i | + | \nabla \times \boldsymbol{v}_i | + \varepsilon_b c_i / h_i},
\end{eqnarray}
where $\varepsilon_b$ is a small value introduced to prevent numerical overflow.
In this paper we set $\varepsilon_b = 0.0001$.
Here, the rotation of velocity is given by
\begin{eqnarray}
\nabla \times \boldsymbol{v}_i = - \sum_j \frac{Y_j}{p_i} \boldsymbol{v}_{ij} \times \nabla W(\boldsymbol{x}_{ij} ; h_i).
\end{eqnarray}

Consequently, the viscosity terms for the equation of motion and the equation of energy are given by
\begin{eqnarray}
\left( m_i \frac{d\boldsymbol{v}_i}{dt} \right)_{\rm AV} & = & - m_i \sum_{j} m_j \frac{1}{2}(F_i + F_j) \Pi_{ij} \frac{1}{2} \left[ \nabla W(\boldsymbol{x}_{ij} ; h_i) + \nabla W(\boldsymbol{x}_{ij} ; h_j) \right],\\
\left( \frac{dU_i}{dt} \right)_{\rm AV} & = & \frac{m_i}{2}\sum_j m_j \frac{1}{2}(F_i + F_j) \Pi_{ij} \boldsymbol{v}_{ij} \cdot \frac{1}{2} \left[ \nabla W(\boldsymbol{x}_{ij} ; h_i) + \nabla W(\boldsymbol{x}_{ij} ; h_j) \right],
\end{eqnarray}
respectively.

One might imagine that the use of the smoothed density in the artificial viscosity would be inconsistent with the formulation of our new SPH.
The artificial viscosity is, however, a mimic of the molecular dissipation, which is not included in the original set of equations for hydrodynamics.
Thus, the choice of the form of the artificial viscosity is independent from the formulation of the SPH.

\subsection{Timestep}
The timesteps for integration are limited by the Courant condition for numerical stability.
The timesteps of the $i$-th particle $dt_i$ is given by
\begin{eqnarray}
dt_i = C_{\rm CFL}\frac{2 h_i}{\max_j v_{ij}^{\rm sig}}.
\end{eqnarray}
We use shared timestep and adopt $dt = \min_{i} dt_i$ as a time step of each step.
Throughout this paper, we adopt $C_{\rm CFL} = 0.3$.

\section{Numerical Tests}
In this section, we report the results of several 1D and 2D tests for non-ideal EOS with the standard SPH and our new SPH.
For both methods, we use the equations with $\nabla h$ term.
Note that our new SPH method reduces to Saitoh \& Makino (2013)'s SPH method, in the case of the ideal gas EOS.
We have confirmed that our new SPH can reproduce the results of Saitoh \& Makino (2013) well when we adopted the EOS of the ideal gas.
Here, we only show the results for non-ideal EOS.

\subsection{Shock tube tests}
The shock tube test is one of the most common test problems.
It is designed to test the ability of numerical method to capture the shock.
We place the initial discontinuity at the origin of the coordinates.
We place equal-mass particles.
The particle separation varies according to the density distribution.
In this section, we introduce the result of 1D shock tube test for non-ideal gas, the Tammann EOS (Ivings et al. 1998), for which the exact solutions exist.
This test was first performed by Wu \& Shen (2008).
The Tammann EOS is suitable for liquid at high pressure.
The initial conditions of this test are as follows:
\begin{eqnarray}
(\rho , p , v) = \left\{
\begin{array}{lll}
(1 , 1000 , 20) & {\rm for} & x < 0,\\
(1 , 1 , 20) & {\rm for} & x \geq 0.
\end{array} \right.
\end{eqnarray}
The Tammann EOS is given by
\begin{eqnarray}
p = (\gamma - 1) \rho u - \gamma p_c,
\end{eqnarray}
where we set $\gamma = 7.15$ and $p_c = 3309$.
The parameter $\eta$ for the smoothing length was set to $2.4$.

\begin{figure}[tbp]
	\begin{minipage}{0.49\textwidth}
		\centering\FigureFile(70mm, 50mm){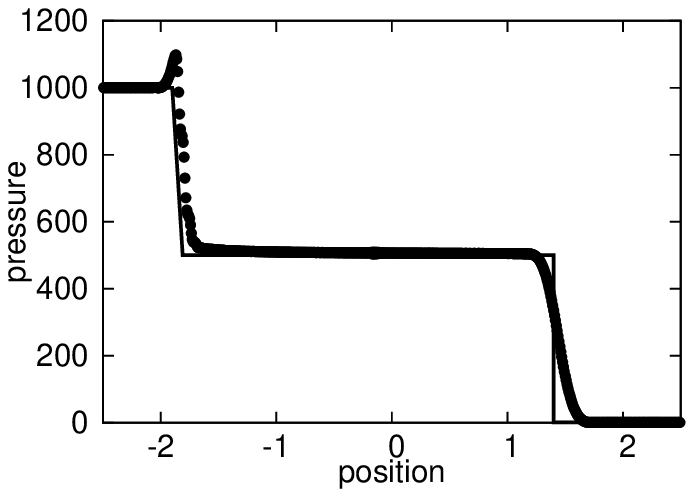}
	\end{minipage}
	\begin{minipage}{0.49\textwidth}
		\centering\FigureFile(70mm, 50mm){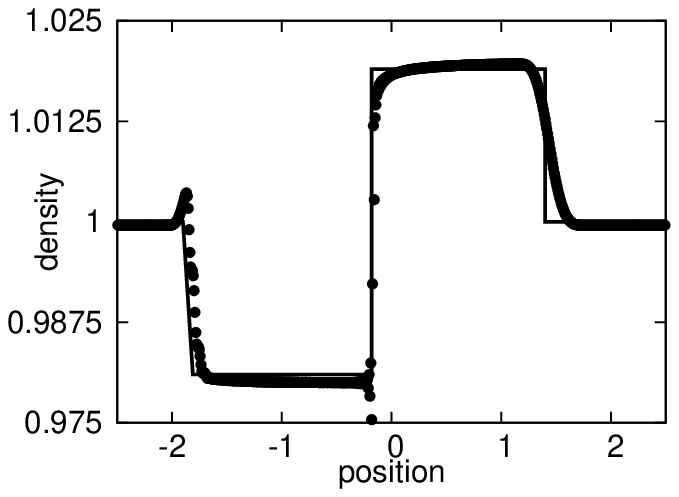}
	\end{minipage}
	\\
	\begin{minipage}{0.49\textwidth}
		\centering\FigureFile(70mm, 50mm){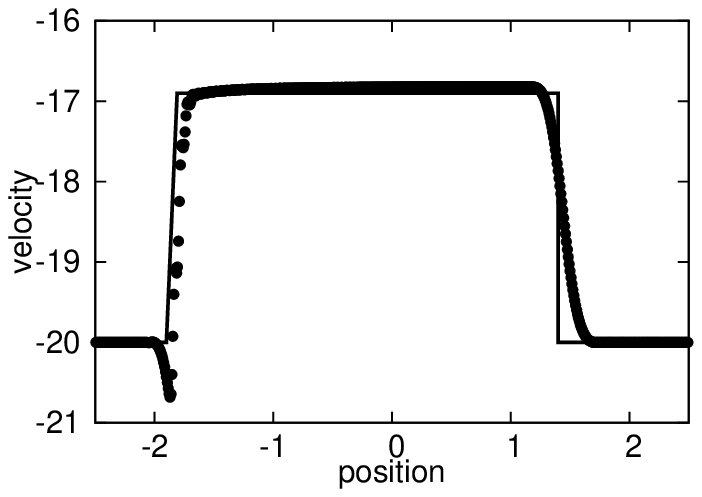}
	\end{minipage}
	\begin{minipage}{0.49\textwidth}
		\centering\FigureFile(70mm, 50mm){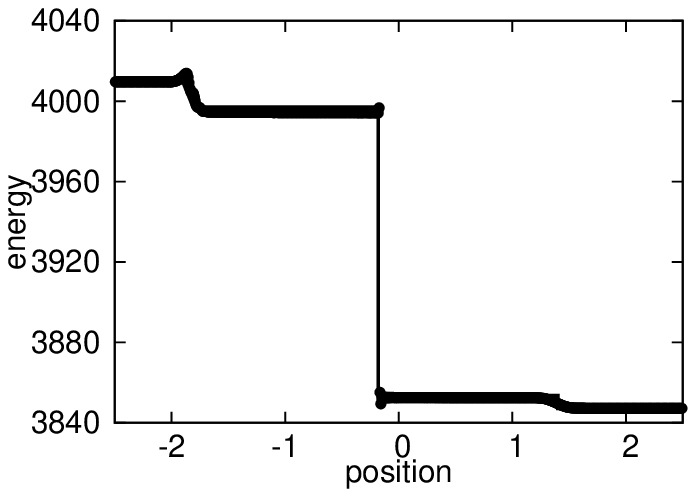}
	\end{minipage}
	\caption{Snapshots from the 1D shock tube test for the Tammann EOS with our new SPH at $t = 0.01$.
        The dots indicate SPH particles and the solid curves represent the exact solution.}
	\label{fig:PF_Water}
\end{figure}

\begin{figure}[tbp]
	\begin{minipage}{0.49\textwidth}
		\centering\FigureFile(70mm, 50mm){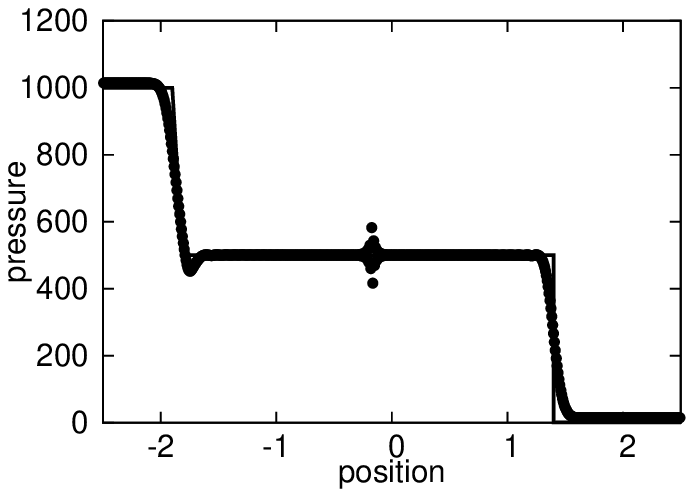}
	\end{minipage}
	\begin{minipage}{0.49\textwidth}
		\centering\FigureFile(70mm, 50mm){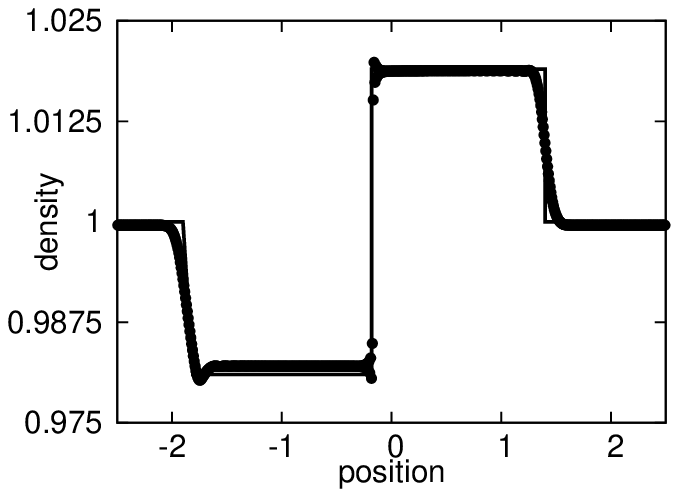}
	\end{minipage}
	\\
	\begin{minipage}{0.49\textwidth}
		\centering\FigureFile(70mm, 50mm){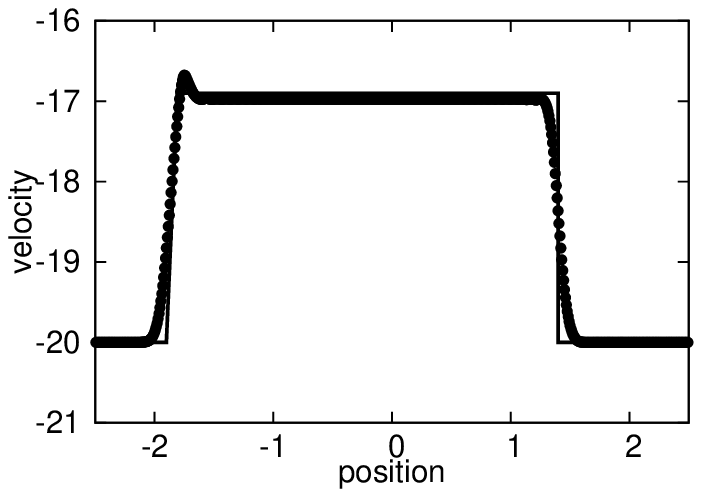}
	\end{minipage}
	\begin{minipage}{0.49\textwidth}
		\centering\FigureFile(70mm, 50mm){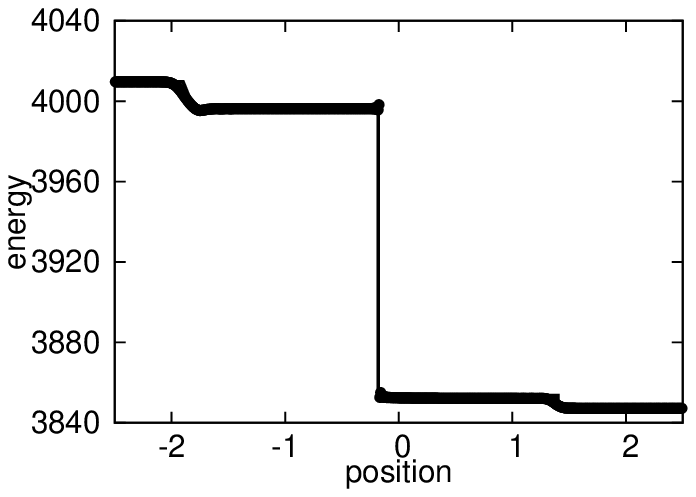}
	\end{minipage}
	\caption{Same as Fig. $\ref{fig:PF_Water}$ but with the standard SPH.
	At the contact discontinuity, a large pressure blip can be seen.
	This causes a suppression of fluid mixing.}
	\label{fig:STD_Water}
\end{figure}

\begin{figure}[tbp]
	\centering\FigureFile(70mm, 50mm){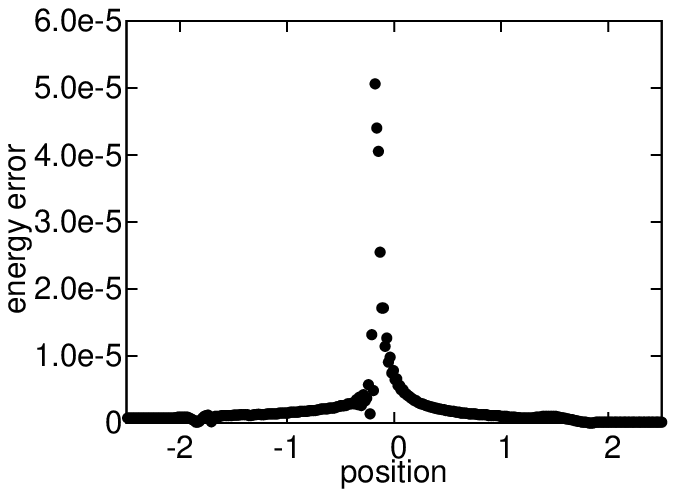}
	\caption{Energy error of the 1D shock tube test for the Tammann EOS at $t = 0.01$.}
	\label{fig:water_energy_error}
\end{figure}

Figure \ref{fig:PF_Water} shows the results of our new SPH at time $t = 0.01$ and figure \ref{fig:STD_Water} shows those of the standard SPH.
At the contact discontinuity, $x \simeq - 0.2$, the standard SPH produces a large pressure blip, whereas our new SPH eliminates this blip.
Our new SPH can handle the contact discontinuity much better, even for non-ideal gas.

Our new method produced somewhat larger overshooting at the front of the rarefaction wave.
In this strong shock test, the pressure is initially strongly discontinuous, while the density is continuous.
As a consequence, with our new SPH, strong overshooting around the contact discontinuity appears in the first several time steps and remains there until the end of simulation.

Figure \ref{fig:water_energy_error} shows the relative error of the specific internal energy for each particle as defined in Eq. (\ref{eq:err2}).
At the contact discontinuity, $x \simeq 0.2$, large errors can be seen.
However, for any particle, the absolute value of the error is less than 0.01\%.

\subsection{Hydrostatic equilibrium tests}
This test clearly shows the ability of a scheme to handle the contact discontinuity.
The similar test has been performed by Saitoh \& Makino (2013) with the ideal gas EOS.
In order to check the ability of our new SPH to the non-ideal EOS, we use the Tillotson EOS (see below), instead of the ideal gas EOS.
We set a high-density region in a low-density ambient, at a pressure equilibrium.
We use a 2D computational domain, $0 \leq x < 1$ and $0 \leq y < 1$.
In both directions, the mirror boundary condition is imposed.
The density is
\begin{eqnarray}
\rho = \left\{
\begin{array}{ccccc}
4 & {\rm for} & 0.25 < x < 0.75 & {\rm and} & 0.25 < y < 0.75,\\
1 & {\rm otherwise}.
\end{array} \right.
\end{eqnarray}
To express the above density distribution, we place equal-mass particles in a uniform grid.
The number of particles in the dense square is 4225 and that in the ambient is 3007, respectively.
The end time is $t = 8$.
Since the system is in the hydrostatic equilibrium, particles should not move.

The Tillotson EOS (Tillotson 1962; Melosh 1989) is one of the most widely used EOS for giant impact simulations (e.g., Benz et al. 1986; Canup \& Asphaug 2001; Genda et al. 2012).
The Tillotson EOS contains 10 parameters, which we should choose to describe given material.
The Tillotson EOS takes three different functional forms depending on the density $\rho$ and the specific internal energy $u$.

(A)condensed ($\rho > \rho_0$) or cold state ($u < u_{\rm iv}$)\\
In this region, the Tillotson EOS is given by the following form:
\begin{eqnarray}
p_{\rm co} = \left( a + \frac{b}{\displaystyle\frac{u}{u_0 \eta^2} + 1} \right) \rho u + A \mu + B \mu^2.
\end{eqnarray}
where $\eta = \rho / \rho_0$ and $\mu = \eta - 1$.

(B)expanded hot state ($\rho < \rho_0$ and $u > u_{\rm cv}$)\\
In this region, the Tillotson EOS is given by the following form:
\begin{eqnarray}
p_{\rm ex} = a \rho u + \left[ \frac{b \rho u}{\displaystyle\frac{u}{u_0 \eta^2} + 1} + A \mu \exp\left\{- \alpha \left( \frac{1}{\eta} - 1\right) \right\} \right] \exp\left\{- \beta \left( \frac{1}{\eta} - 1\right)^2 \right\}.
\end{eqnarray}

(C)intermediate region ($u_{\rm iv} < u < u_{\rm cv}$ and $\rho < \rho_0$)\\
In this region, a smooth transition between above two states occurs.
Thus, as Benz et al. (1986) did, we interpolated the pressure by using $p_{\rm co}$ and $p_{\rm ex}$;
\begin{eqnarray}
p_{\rm tr} = \frac{(u - u_{\rm iv}) p_{\rm ex} + (u_{\rm cv} - u)p_{\rm co}}{u_{\rm cv} - u_{\rm iv}}.
\end{eqnarray}
Here, $\rho_0, u_0, a, b, A, B, u_{\rm cv}, u_{\rm iv}, \alpha$ and $\beta$ are material parameters.
In this paper, we use the values for granite: $\rho_0 = 2680~{\rm kg/m}^3, u_0 = 16~{\rm MJ/kg}, a = 0.5, b = 1.3, A = 18~{\rm GPa}, B = 18 ~{\rm GPa}, u_{\rm cv} = 18~{\rm MJ/kg}, u_{\rm iv} = 3.5~{\rm MJ/kg}, \alpha = 5, \beta = 5$.
We set the density unit $\rho_0$, unit specific energy $u_0$ and unit pressure $\rho_0 u_0$.

\begin{figure}[tbp]
	\begin{minipage}{0.175\textwidth}
		\centering\FigureFile(30mm, 30mm){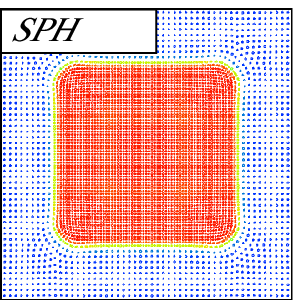}
	\end{minipage}
	\begin{minipage}{0.175\textwidth}
		\centering\FigureFile(30mm, 30mm){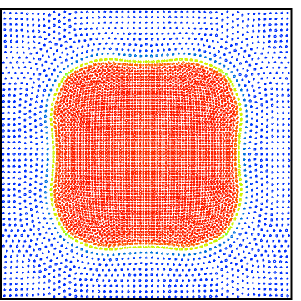}
	\end{minipage}
	\begin{minipage}{0.175\textwidth}
		\centering\FigureFile(30mm, 30mm){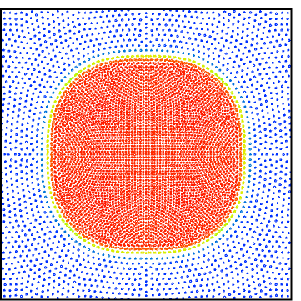}
	\end{minipage}
	\begin{minipage}{0.175\textwidth}
		\centering\FigureFile(30mm, 30mm){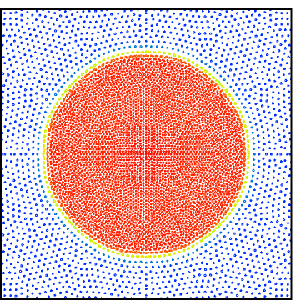}
	\end{minipage}
	\begin{minipage}{0.175\textwidth}
		\centering\FigureFile(30mm, 30mm){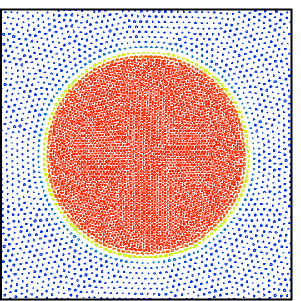}
	\end{minipage}
	\\
	\begin{minipage}{0.175\textwidth}
		\centering\FigureFile(30mm, 30mm){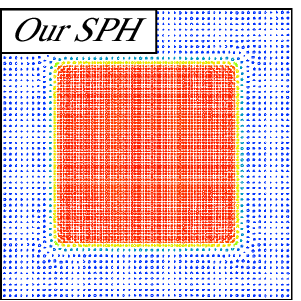}
	\end{minipage}
	\begin{minipage}{0.175\textwidth}
		\centering\FigureFile(30mm, 30mm){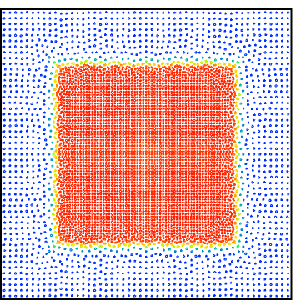}
	\end{minipage}
	\begin{minipage}{0.175\textwidth}
		\centering\FigureFile(30mm, 30mm){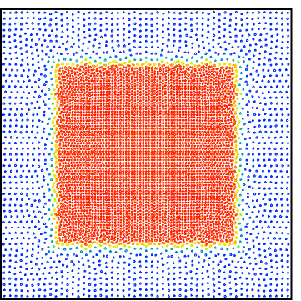}
	\end{minipage}
	\begin{minipage}{0.175\textwidth}
		\centering\FigureFile(30mm, 30mm){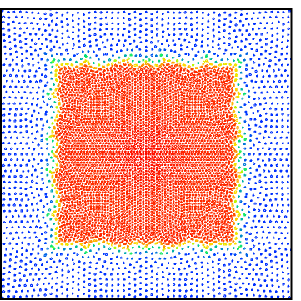}
	\end{minipage}
	\begin{minipage}{0.175\textwidth}
		\centering\FigureFile(30mm, 30mm){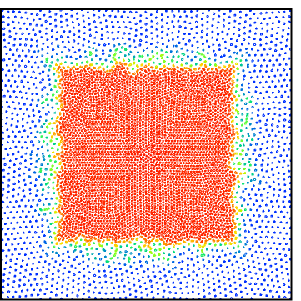}
	\end{minipage}
	\\
	\centering\FigureFile(152.4mm, 10.16mm){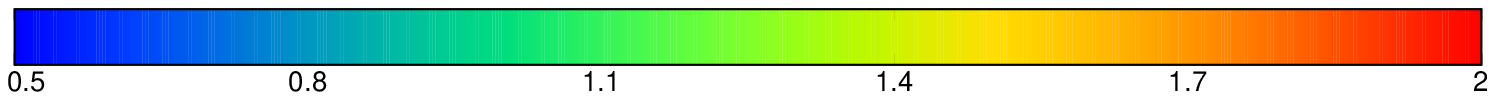}
	\caption{Snapshots from the hydrostatic equilibrium tests at $t = 0.1, 0.3, 0.5, 1$ and $8$ from left column, respectively.
	The top panels show the results of the standard SPH and the bottom panels show those of our new SPH.
	The dots represent the positions of SPH particles.
	The color code of the density is given at the bottom.}
	\label{fig:Sq}
\end{figure}
Figure \ref{fig:Sq} shows the results of this tests for the standard SPH and our new SPH.
The difference between two scheme is clear.
With the standard SPH, although the pressures of each particle are initially equal, the high-density domain becomes a circle at $t = 8$.
The reason why such an unphysical transform occurs is explained Saitoh \& Makino (2013).

In contrast, with our new SPH, the high-density domain keeps its original shape, except some local rearrangement near the boundary of two fluids.
Our new SPH removes the unphysical surface tension completely, even for non-ideal gas.

\subsection{KHI tests}
KHI is one of the most fundamental test problems for the ability of numerical methods to handle hydrodynamical instability.
Initially, two layers in pressure equilibrium has the different density and move to opposite direction to each other.

We perform the KHI test for the Tillotson EOS.
We use a 2D computational domain, $-0.5 < x \leq 0.5$ and $-0.25 < y \leq 0.25$.
The periodic boundary conditions are imposed in the $x$-direction and the mirror boundary condition is imposed in the $y$-direction.
We set the density as follows:
\begin{eqnarray}
\rho = \left\{
\begin{array}{ccccc}
\rho_l & = & 1 & {\rm for} & y > 0,\\
\rho_h & = & 2 & {\rm for} & y \leq 0,
\end{array} \right.
\end{eqnarray}
where $\rho_l$ and $\rho_h$ are the density of the low-density region and that of the high-density region, respectively.
The shear velocity is set up in the $x$-direction.
We set $v_{x,h} = 0.5$ for the high density region and $v_{x,l} = - 0.5$ for the low density region, respectively.
As a seed of the instability, small perturbation is added to the particles around the initial contact discontinuity:
\begin{eqnarray}
v_y = \Delta v_y \sin \left( \displaystyle\frac{2 \pi x}{\lambda} \right) & \quad {\rm for} \quad |y| < 0.025.
\end{eqnarray}
Here, $\Delta v_y$ and $\lambda$ are the amplitude and wavelength of the initial perturbation, respectively.
We set $\Delta v_y = 0.025$ and $\lambda = 1/6$.
Thus, six vortex rolls are expected to be developed in the computational domain.
The growth time scale of the KHI is
\begin{eqnarray}
\tau_{\rm KH} = \frac{\lambda(\rho_l + \rho_h)}{\sqrt{\rho_l \rho_h} |v_{x,l} - v_{x,h}|}.
\end{eqnarray}
For our test case, $\tau_{\rm KH} \simeq 0.35$.
In each region, we place the equal-mass particles uniformly in a lattice.
The particle separation in the low-density region is set to $1/512$.
The equilibrium pressure is set to $3.5$.

\begin{figure}[tbp]
	\begin{minipage}{0.49\textwidth}
		\centering\FigureFile(66mm, 33mm){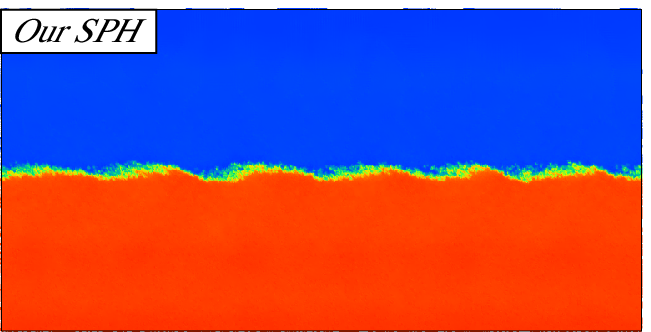}
	\end{minipage}
	\begin{minipage}{0.49\textwidth}
		\centering\FigureFile(66mm, 33mm){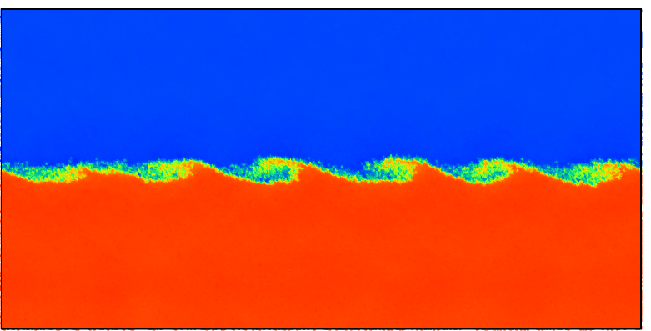}
	\end{minipage}
\\
	\begin{minipage}{0.49\textwidth}
		\centering\FigureFile(66mm, 33mm){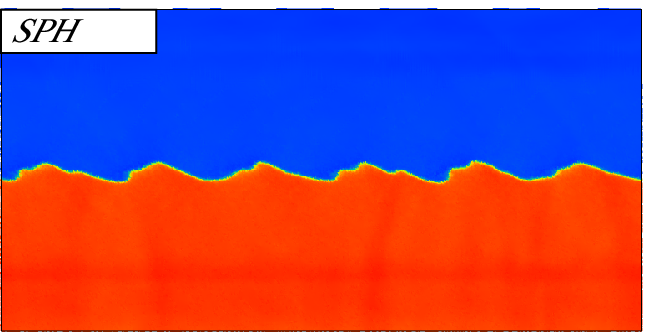}
	\end{minipage}
	\begin{minipage}{0.49\textwidth}
		\centering\FigureFile(66mm, 33mm){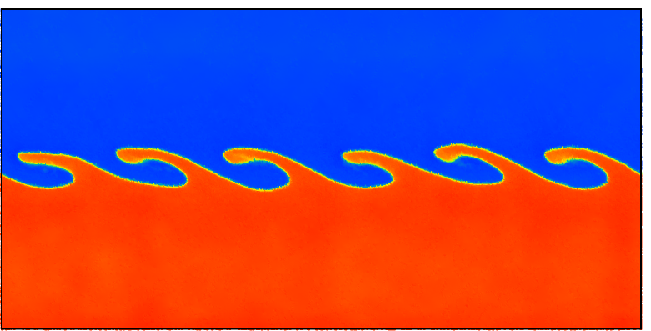}
	\end{minipage}
	\\
	\centering\FigureFile(152.4mm, 10.16mm){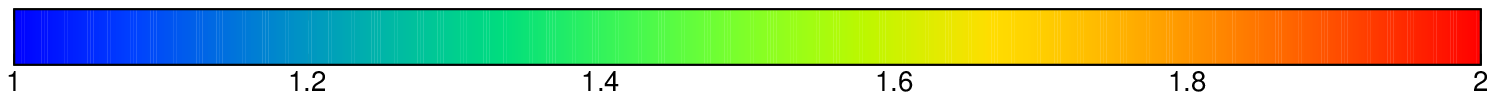}
	\caption{Density distributions from the KHI tests at $t = 1.0~\tau_{\rm KH}$ (left column) and $2.0~\tau_{\rm KH}$ (right column), respectively.
		The upper row is for our new SPH whereas the lower row is for the standard SPH.
		The color code is given at the bottom.
		Density is normalised to $\rho_0$.
	}
	\label{fig:KH_t}
\end{figure}

\begin{figure}[tbp]
	\begin{minipage}{0.49\textwidth}
		\centering\FigureFile(70mm, 70mm){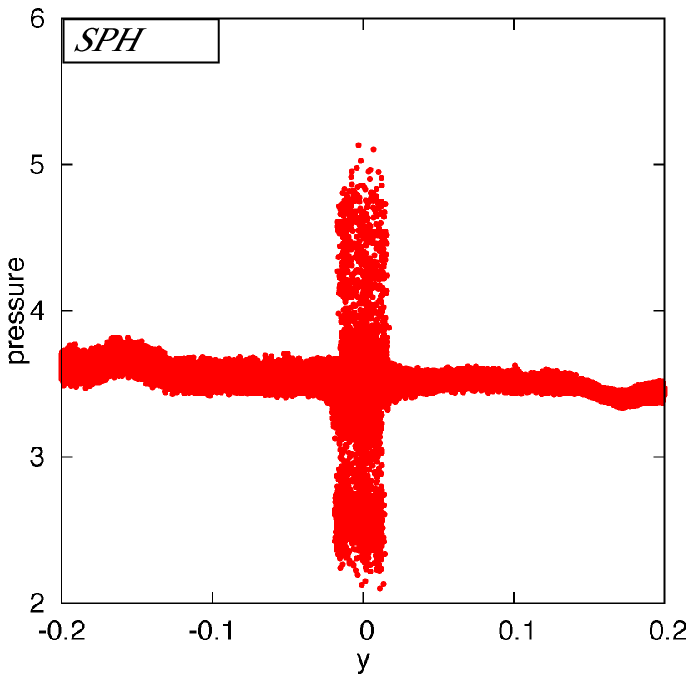}
	\end{minipage}
	\begin{minipage}{0.49\textwidth}
		\centering\FigureFile(70mm, 70mm){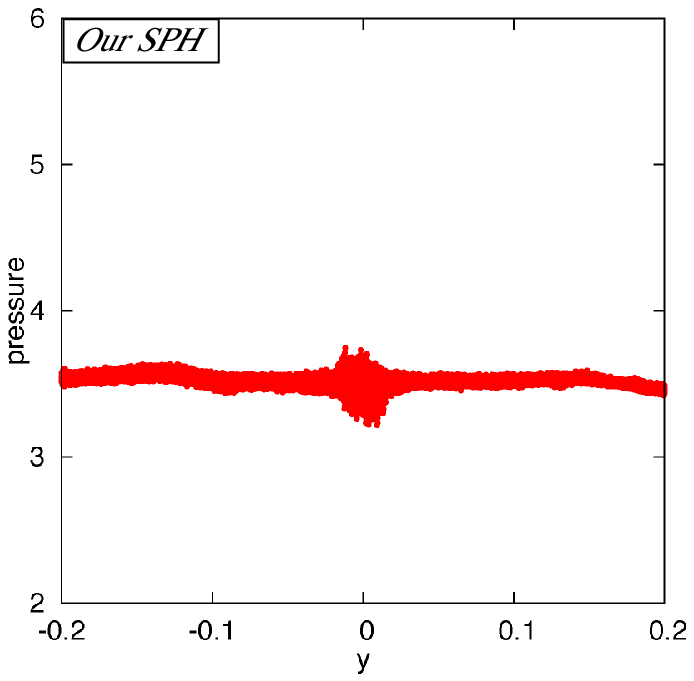}
	\end{minipage}
	\caption{Pressure distribution from the KHI test along the $y$-axis at $t = 1.0~\tau_{\rm KH}$.
		The left panel shows the result of the standard SPH, while the right panel shows that of our new SPH.
		Pressure is normalised to $\rho_0 u_0$.
	}
	\label{fig:KH_t_side}
\end{figure}

\begin{figure}[tbp]
	\begin{minipage}{0.49\textwidth}
		\centering\FigureFile(70mm, 70mm){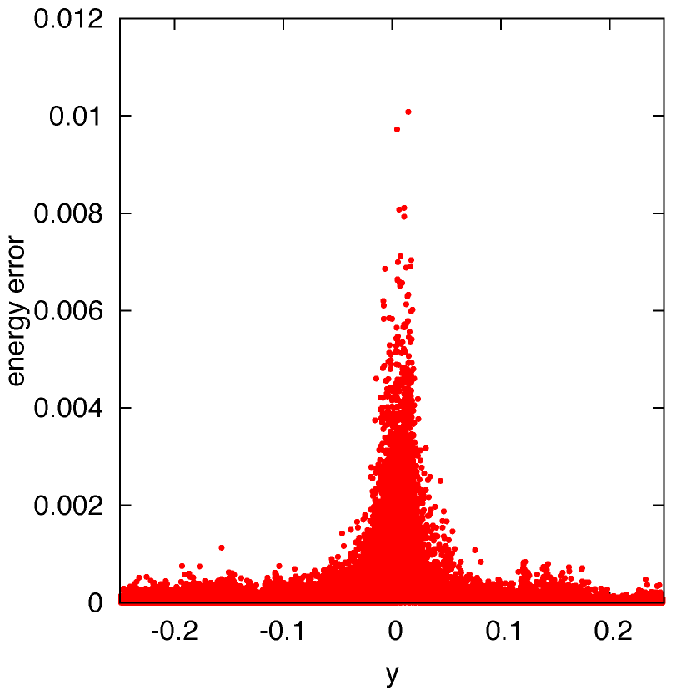}
		\caption{The distribution of the error of the specific internal energy of the KHI test along the $y$-axis at $t = 1.0~\tau_{\rm KH}$ with one iteration.}
		\label{fig:KH_t_energy_error}
	\end{minipage}
	\begin{minipage}{0.49\textwidth}
		\centering\FigureFile(70mm, 50mm){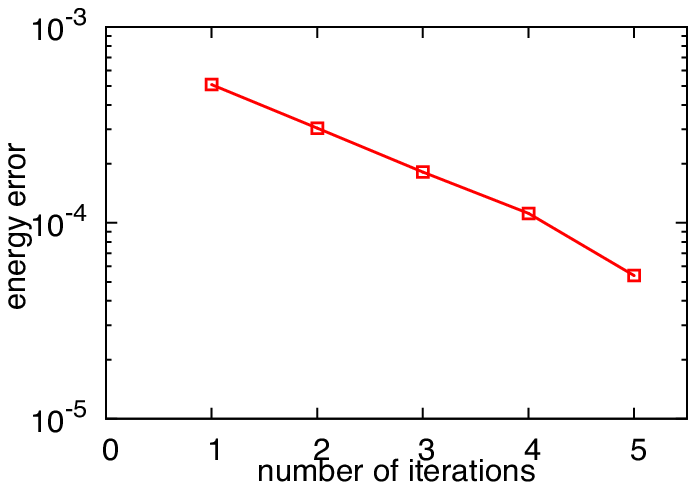}
		\caption{The averaged error of the specific internal energy versus number of iterations for our new SPH.}
		\label{fig:KH_t_times_vs_error}
	\end{minipage}
\end{figure}

Figure $\ref{fig:KH_t}$ shows the density distributions at times $t = 1.0~\tau_{\rm KH}$ and $2.0~\tau_{\rm KH}$.
The upper row is the results of our new SPH, and the lower row is those of the standard SPH.
There is an obvious difference between the two results and our new SPH gives far better results compared to that of the standard SPH.
With the standard SPH, perturbations grow until $1.0~\tau_{\rm KH}$.
However, the unphysical surface tension inhibits the growth of the vortex rolls.
The dense fluid is stretched.
As a consequence, the standard SPH produces ``blobs'' of dense fluid (see Figure 7 in Price 2008; Figure 7 in Saitoh \& Makino 2013).
The mixing between the two layers is completely suppressed.
On the other hand, our new SPH shows very good result.
At $t = 2.0~\tau_{\rm KH}$, six vortex rolls are clearly visible.

Figure $\ref{fig:KH_t_side}$ shows the pressure distribution along $y$-axis at $t = 1.0~\tau_{\rm KH}$.
The left panel shows the result of the standard SPH while the right panel shows that of our new SPH.
With the standard SPH, there is a large pressure jump at the contact interface, $y \simeq 0$.
With our new SPH, on the other hand, the pressure jump is much smaller.
Our new SPH eliminated the unphysical surface tension, even for non-ideal gas.
Thus, the growth of the KHI is not suppressed.

Figure $\ref{fig:KH_t_energy_error}$ shows the distribution of the error of specific internal energy for each particle as defined in Eq. ($\ref{eq:err2}$).
One iteration of the pressure summation loop is done.
At the contact interface $y \simeq 0$, particles have larger errors.
However, even the largest value, the error is about 1\%.
Figure $\ref{fig:KH_t_times_vs_error}$ shows that the averaged of error of the specific internal energy, as defined in Eq. (\ref{eq:err}), decreases as the number of iterations of the pressure summation loop of our new SPH increases.
For all cases, the averaged error of the specific internal energy is less than 0.1\%, and the error becomes smaller by a factor of two after each iteration.

\section{Discussion and Summary}
\subsection{Treatment of mixing}
In real fluid, mixing takes place due to the physical dissipation, namely, the random motion of molecules.
Thus, if we had an infinite number of particle, the mixing would not take place at all and we could resolve infinitely small vortices, as far as we do not include any physical diffusion term.
For the test of KHI, however, our new SPH produces somewhat noisy contact interface between two fluids.
Therefore, one might think that other schemes, such as the AC term, are better than our new SPH.
However, the noisy interface appears because we have finite number of particles and is at least partly due to the KHI itself at high-wavenumber, which is physically there.
Thus, we argue that the noisy interface is not problematic.

It is worth noting that, in the case of the jump in the chemical composition, the standard AC term is insufficient and it is necessary to introduce artificial chemical diffusion term.
Our scheme can handle any kind of discontinuity without any diffusion term. 
Of course, to express the fluid mixing in the sub-resolution scale appropriately, we should introduce the turbulent diffusion term (Wadsley et al. 2008).

\subsection{Summary}
The SPH method is a powerful numerical tool for astrophysical and planetlogical problems.
However, due to the requirement of the differentiability of density, the standard SPH has a problem in describing multi-phase flows and mixing.
In this paper, we describe an alternative formulation of SPH in which the pressure is used as the basis of the smoothing instead of the density.
In our formulation, we do not assume the differentiability of the density, but assume that of the pressure.
As a result, our new formulation shows great improvement in the treatment of contact discontinuity and hydrodynamical instabilities.
Our new SPH can handle problems in which mixing takes place.
Our new SPH is natural extension of that of Saitoh \& Makino (2013).
With our new SPH, the shock tube, the hydrostatic equilibrium test and the KHI test show good results for non-ideal gas.
It is easy and straightforward to modify existing SPH to our new method.
In addition, our new SPH does not introduce any additional dissipation term and does not break conservation properties.
The increase of the calculation cost is small.

Our new SPH can be easily incorporated with other improvements, for example inviscid SPH (Morris \& Monaghan 1997; Cullen \& Dehnen 2010) and higher-order dissipation switch (Read \& Hayfield 2012).

One important application of our new SPH is the giant impact simulations, where the instabilities and mixing at the boundaries of different materials might play important roles in.
We are currently working to apply our new SPH to the giant impact simulations.
The results will be reported in the forthcoming paper.
Of course, our new SPH can be applied to a variety other astrophysical and planetlogical problems.

\bigskip
The authors thank the anonymous referees for giving us helpful comments on the manuscript.
This work is supported by a grant for the Global COE Program, `From the Earth to ``Earths"', MEXT, Japan.
It was also supported in part by a Grant-in-Aid for Scientific Research (21244020) and Strategic Programs for Innovative Research of the Ministry of Education, Culture, Sports, Science and Technology (SPIRE).

\end{document}